# Emergent functions of noise-driven spontaneous activity: Homeostatic maintenance of criticality and memory consolidation

**Narumitsu Ikeda[1], Dai Akita[1], and Hirokazu Takahashi[1,2*]**

[1]Department of Mechano-Informatics, Graduate School of Information Science and Technology, The University of Tokyo, Tokyo, Japan

[2]International Research Center for Neurointelligence (WPI-IRCN), UTIAS, The University of Tokyo, Tokyo, Japan

*** Correspondence:**
Hirokazu Takahashi
takahashi@i.t.u-tokyo.ac.jp

**Keywords: Spiking neural network (SNN), self-organized criticality, EI balance, fading memory, reservoir computing, dissociated neuronal culture**

**Abstract**

Unlike digital computers, the brain exhibits spontaneous activity even during complete rest, despite the evolutionary pressure for energy efficiency. Inspired by the critical brain hypothesis, which proposes that the brain operates optimally near a critical point of phase transition in the dynamics of neural networks to improve computational efficiency, we postulate that spontaneous activity plays a homeostatic role in the development and maintenance of criticality. Criticality in the brain is associated with the balance between excitatory and inhibitory synaptic inputs (EI balance), which is essential for maintaining neural computation performance. Here, we hypothesize that both criticality and EI balance are stabilized by appropriate noise levels and spike-timing-dependent plasticity (STDP) windows. Using spiking neural network (SNN) simulations and in vitro experiments with dissociated neuronal cultures, we demonstrated that while repetitive stimuli transiently disrupt both criticality and EI balance, spontaneous activity can develop and maintain these properties and prolong the fading memory of past stimuli. Our findings suggest that the brain may achieve self-optimization and memory consolidation as emergent functions of noise-driven spontaneous activity. This noise-harnessing mechanism provides insights for designing energy-efficient neural networks, and suggest a potential link between the emergent function of spontaneous activity and sleep function in maintaining homeostasis and consolidating memory.

## 1 Introduction

The brain differs significantly from a digital computer in that it displays spontaneous activity, even in a state of complete rest. Why does the brain not eliminate spontaneous activity even under strong evolutionary selection pressure for an energy-efficient system? As a possible explanation, we propose that spontaneous activity serves a crucial function: helping the brain reach and maintain a critical state. This is based on the critical brain hypothesis, which states that neural networks in the brain operate near a critical point of phase transition between ordered (inactive) and chaotic (overactive) states to achieve computational efficiency (1-5). Supporting this hypothesis is evidence of self-organized criticality (SoC) in the brain, characterized by specific patterns of neural activity known as “neuronal avalanche,” which display spatial and temporal self-similarity with power-law correlations (2, 6-8). This SoC is likely to enable a balance between excitatory and inhibitory synaptic inputs (EI balance) (9-12), which is essential for the generation of temporally irregular spike patterns (13-15) and the maintenance of neural computation performance (16-20).

In the critical state, the population of active neurons neither decays nor grows over time, in contrast to the subcritical and supercritical states(5, 21, 22). This stability enables critical neural networks to maintain a longer fading memory –preserving influences from past inputs and allowing interaction with network dynamics (23-27)– and achieve higher computational performance than non-critical networks in information transmission(6, 9), encoding and storage (7, 28-31). However, questions remain regarding how SoC persists amidst the inevitable disruptions caused by learning and plasticity.

Dissociated neuronal cultures –simplified neural networks created in vitro– have long served as experimental models for understanding how SoC emerges in neural systems (2, 6, 21, 32, 33), whereas repeated external stimuli induce plasticity in neuronal cultures in an activity-dependent manner and modify the network topology(34-37). Plasticity can strengthen the memory trace of repeated inputs in the neuronal network, but in turn disrupts the critical state. To our knowledge, no study to date has combined these two important research topics in dissociated neuronal culture.

Our recent work demonstrated that SoC development in dissociated neuronal cultures can be reproduced by simulations of spiking neural networks (SNNs) (11, 33). In these simulations, we found that appropriate noise levels and time windows of spike-timing-dependent plasticity (STDP) were sufficient for developing and maintaining both the critical state and EI balance in SNNs (11). This finding is consistent with the emerging notion that criticality and EI balance serve as homeostatic set points in cortical neuronal dynamics (38-42). Given that criticality and EI balance are regulated by noise and STDP, we hypothesized that this homeostatic mechanism also enables neural networks to better encode traces of past input stimuli.

In this study, we tested our hypothesis by using both SNNs and dissociated neuronal cultures. Our hypothesis predicts that repeated stimuli to the network will induce stimulus-specific plasticity in a Hebbian fashion, thereby disrupting the critical state. However, spontaneous activity following this period of repetitive stimulation is expected to restore the critical state with prolonged stimulus-specific fading memory. Our results confirmed these predictions: in both the SNN and neuronal cultures, repeated stimulation initially disrupted the critical state. However, spontaneous activity soon restored this balance, with the criticality index of the SNN returning to baseline within two hours. In addition, both systems showed prolonged, stimulus-specific fading memory after this recovery period, with dissociated cultures displaying heightened stimulus-specific responses following spontaneous activity.

A better understanding of the functional role of spontaneous activity may inform the development of neuromorphic computation that not only tolerates, but also utilizes noise to achieve system self-optimization. From a physiological perspective, our findings suggest a potential link between the emergent function of spontaneous activity and sleep function in homeostatic maintenance and memory consolidation.

## 2 Materials and methods

### 2.1 Neurons in Spiking Neural Network (SNN)

In accordance with Stepp et al (40) and our previous study (11), we used an SNN model with LIF dynamics and escape noise (43). The membrane potential $v$ without noise is given by

$$\tau_m \frac{dv}{dt} = (v_{rest} - v)g_{rest} + (E_{exc} - v)g_{exc} + (E_{inh} - v)g_{inh}, \qquad (1)$$

where $\tau_m$ is the membrane time constant, $v_{rest}$ is the resting potential, and $E$ and $g$ are the reversal potentials and sum of the conductance for the excitatory synapse (*exc*), inhibitory synapse (*inh*), and resting state (*rest*), respectively. The equation was normalized by $g_{rest}$, i.e., $g_{rest}$=1, and the parameters in Table 1 were used in the simulation. The membrane potential is determined deterministically according to (1) until the neuron fires. According to the escape noise model, the instantaneous firing probability of a neuron is given by

$$\rho(t) = \frac{1}{\tau} \exp\left(\frac{v(t) - v_{th}}{b}\right), \qquad (2)$$

where $\tau$ and $b$ are parameters and $v_{rest}$ is the firing threshold. With time discretization by time step $\Delta t$, the probability of firing within a given time step is given by

$$Prob(spike\ in\ [t, t + \Delta t]) = 1 - \exp\left(-\Delta t \rho\left(v(t)\right)\right). \qquad (3)$$

With a first-order approximation, an upper limit is set as follows

$$Prob(spike\ in\ [t, t + \Delta t]) = \min\left(\frac{\Delta t}{\tau} \exp\left(\frac{v(t) - v_{th}}{b}\right), 1\right), \qquad (4)$$

$$\frac{\Delta t}{\tau} = f_{rest} \Delta t \exp\left(-\frac{v_{rest} - v_{th}}{b}\right). \qquad (5)$$

With the parameters as shown in Table 1, $\Delta t/\tau$~0.00594 ($\tau$~16.8), which is close to the corresponding parameter of the pyramidal neuron at layer 5 in the rat somatosensory cortex, i.e., 0.1 ms/19 ms~0.00526.

When a neuron fires, $v$ is reset to $v_{rest}$ with refractory periods $T_{exc}^{ref}$ and $T_{inh}^{ref}$ for the excitatory and inhibitory neurons, respectively.

A synaptic input is given to postsynaptic neurons with a delay of 1.5 ms for excitatory-to-excitatory synapses and 0.8 ms for others. The excitatory and inhibitory conductances ($g_{exc}$ and $g_{inh}$) are given by

$$\tau_{AMPA} \frac{dg_{exc}}{dt} = -g_{exc}, \qquad (6)$$

$$\tau_{GABA} \frac{dg_{inh}}{dt} = -g_{inh}. \qquad (7)$$

When a presynaptic spike occurs, conductance is updated as follows:

$$g_{exc} \leftarrow g_{exc} + U\ x\ w\ g_{exc}^{max}, \qquad (8)$$

$$g_{inh} \leftarrow g_{inh} + U\ x\ w\ g_{inh}^{max}, \qquad (9)$$

where U is a parameter corresponding to the fraction of resources used in a spike, $w$ is the synaptic weight, and $g_{exc}^{max}$ and $g_{inh}^{max}$ are the maximum synaptic conductances of the excitatory and inhibitory neurons, respectively. Variable $x$ corresponds to the available resource in a presynaptic neuron, which recovers to 1 with a time constant $\tau_{rec}$:

$$\tau_{rec}\frac{dx}{dt} = 1 - x. \qquad (10)$$

The variables $x$ and $w$ are modulated by short-term plasticity (STP) and STDP on a spike event, as described in the next section.

## 2.2 Synaptic plasticity in SNN

In the event of presynaptic firing, the STP updates $x$ as follows:

$$x \leftarrow x - U\,x. \qquad (11)$$

STDP updates $w$ on both pre- and postsynaptic spike events. When a presynaptic spike is generated at time $t_{pre}$,

$$w \leftarrow \mathrm{Clip}\left( w + \sum_{t_{post} \in T_{post}(t_{pre})} F\left(t_{post} - t_{pre}\right), 0, 1 \right), \qquad (12)$$

where $T_{post}(t_{pre})$ is the set of postsynaptic spike times before $t_{pre}$, and $F(t)$ is a synapse-type-dependent function, and Clip is a clipping function given by

$$\mathrm{Clip}(w, a, b) = \min(\max(w, a), b). \qquad (13)$$

When a postsynaptic spike is generated at $t_{post}$,

$$w \leftarrow \mathrm{Clip}\left( w + \sum_{t_{pre} \in T_{pre}(t_{post})} F\left(t_{post} - t_{pre}\right), 0, 1 \right), \qquad (14)$$

where $T_{pret}(t_{post})$ is the set of postsynaptic spike times before $t_{post}$. When a presynaptic neuron is excitatory (E-STDP), $F(t)$ is given by

$$F_E(t) = \begin{cases} A_E \exp\left(-\frac{t}{\tau_E}\right), & t \geq 0, \\ -A_E\,\beta_E \exp\left(-\frac{t}{\tau_E}\right), & t < 0. \end{cases} \qquad (15)$$

When a presynaptic neuron is inhibitory (I-STDP), $F(t)$ is given by

$$F_E(t) = \frac{A_I}{1 - \frac{\tau_{I1}}{\tau_{I2}}\beta_I}\left( \exp\left(-\frac{|t|}{\tau_{I1}}\right) - \frac{\tau_{I1}}{\tau_{I2}}\beta_I \exp\left(-\frac{|t|}{\tau_{I2}}\right)\right). \qquad (16)$$

The parameters $A_E$ and $A_I$ (>0) regulate the intensity of plasticity. $\tau_E$, $\tau_{I1}$, and $\tau_{I2}$ are time constants assuming that

$$\frac{\tau_{I1}}{\tau_{I2}} < \min\left(\frac{1}{\beta_I}, 1\right). \qquad (17)$$

Parameters $\beta_E$ and $\beta_I$ determine the balance between potentiation and depression (Fig. 1A).

### 2.3 Implementation of SNN

Our simulation was performed using a Brian 2 100 neural simulator(44). The parameters were based on previous studies (24, 40, 45) and are summarized in Table 1.

The SNN consisted of 80 excitatory neurons and 20 inhibitory neurons that were fully connected with an initial weight of 0, as in the dissociated neuronal cultures. Neurons in the initial state are activated only by noise and the SNN evolves through synaptic plasticity. For each condition, the simulations were run 30 times with different random seeds of $v$ in the initial state.

### 2.4 Stimulation in SNN

For stimulation, 4 different spatial stimulus patterns were created for each SNN. In each stimulus pattern, the membrane potentials of six different excitatory neurons were added by 30 mV. Because the resting membrane potential and firing threshold were -74 mV and -54 mV, respectively, neurons with membrane potentials near or above the resting membrane potential generated a spike upon stimulation.

Stimulus patterns were applied repeatedly to induce stimulus-specific plasticity in each SNN. Each stimulus pattern was applied 900 times at 1 Hz, for a total of 3600 stimuli. Stimulation patterns were presented in random order. After repeated stimulation, SNNs were allowed to spontaneously run for 12 h. Spontaneous activity was measured every hour for 10 min to characterize the neural avalanche and monitor synaptic weights, membrane potentials, and synaptic conductance.

### 2.5 Neural Avalanche Analysis

Based on our previous studies (11, 33) neural avalanches are characterized by 10-min spontaneous activities. The spike times of all the neurons were obtained as a single series $\{t_1, t_2, \ldots\}$. A neural avalanche is defined as a separate set of spikes $\{t_i, \ldots, t_{i+n-1}\}$ (for any $j \in [i+1, i+n-1], t_j - t_{j-1} < \Delta t$), where $\Delta t$ is the average spike interval of the spike time series and the avalanche size $n$ is defined as the number of spikes in the avalanche.

The criticality of the SNN was evaluated using the index ΔCr (11), which was modified from the index $\Delta p$ by Tetzlaff et al. (46). Let $s$ be the size of the neural avalanche and $p_{fit}(s)$ be the power-law distribution estimated from linear regression on the log-log plot of the empirical neural avalanche distribution $p_{emp}(s)$, $\Delta p$ is given by

$$\Delta p = \sum_{s=s_{min}}^{s_{max}} p_{emp}(s) - p_{fit}(s), \qquad (18)$$

where $s_{min}$ and $s_{max}$ are the minimum and maximum avalanche size, respectively. The parameter $s_{max}$ was set to 100, that is, the total number of neurons, for the SNN simulations, and the number of recording electrodes for the in vitro experiments of neuronal dissociated cultures. The parameter $s_{min}$ is determined to minimize the sum of the squared errors of the linear fit because small avalanche sizes often deviate from the power law fitting (47). ΔCr was used instead of $\Delta p$ to avoid false criticalityjudgment by $\Delta p$ when the deviation from the power-law distribution was large but symmetrical. Considering both the upper and lower deviation, i.e.,

$$A_{upper} = \sum_{s=s_{min}}^{s_{max}} \max\left(p_{emp}(s) - p_{fit}(s), 0\right), \qquad (19)$$

$$A_{lower} = \sum_{s=s_{min}}^{s_{max}} \min\left(p_{emp}(s) - p_{fit}(s), 0\right), \qquad (20)$$

ΔCr is given by

$$\Delta \mathrm{Cr} = \begin{cases} A_{upper}, & \left|A_{upper}\right| \geq \left|A_{lower}\right|, \\ A_{lower}, & \left|A_{upper}\right| < \left|A_{lower}\right|. \end{cases} \qquad (20)$$

The SNN is considered subcritical, critical, and supercritical when ΔCr<0, ΔCr≈0 and ΔCr>0, respectively. For each SNN, the mean and standard deviation of ΔCr were derived from 30 initial seeds.

## 2.6 EI Balance Analysis

To examine how excitatory and inhibitory synaptic inputs to each neuron are coupled in time, synaptic currents from excitatory (E) and inhibitory (I) neurons, i.e., $(E_{exc} - v)g_{exc}$ and $(E_{inh} - v)g_{inh}$ in eq. (1), were measured for 2 s during spontaneous activity, and their cross-correlation was quantified to evaluate the EI balance. The mean synaptic currents from excitatory and inhibitory neurons were averaged and the ratio of inhibitory input to excitatory input (I/E) was calculated.

## 2.7 Fading Memory Analysis

To quantify the hidden memory in a neural network, a decoder of sparse logistic regression (SLR) was constructed to predict which stimulus out of the four test stimuli was used from a spatial firing pattern within a given bin. Evoked responses to each of the four test stimuli were obtained in 40 trials, each with a different random seed to generate stochastic firing in equation (4). As evoked responses also depend on the membrane potential in each neuron and the conductance in each synapse at the moment of stimulation, a 10-s free run with each random seed was given before stimulation. The stimulus-evoked firing rate of each neuron was then obtained to construct a 100-dimensional firing-rate vector for each bin. Using the firing rate vectors for each bin, the SLR classification accuracy was quantified using 10-segment stratified shuffle cross-validation. Friedman's chi-square test was used to test a significant difference in classification accuracies before and after repeated stimuli in each bin.

The classification accuracy was close to 1 immediately after the stimulus onset and gradually decreased over time (Fig. 2). The decay time for each SNN was defined as the time after the stimulus when the accuracy first became the chance rate. A binomial test in the leave-one-out cross-validation was used to determine whether the number of correct responses was significantly higher than the chance rate for each bin ($p<0.05$). The Wilcoxon signed-rank test was used to determine whether the decay time changed before and after the repetitive stimulation session.

The classification accuracies varied depending on the bin width used to calculate the firing rate, but the accuracies and decay times were similar for bin widths between 10 and 50 ms (Fig. S3). Therefore, we used a 20-ms bin width in the main analyses. The classification accuracy and decay time were also characterized as a function of the total number of repetitive stimuli, ranging from 400 to 3600. Each stimulus pattern was presented equally frequently.

## 2.8 Culture Preparation

The experimental protocol was approved by the Committee on the Ethics of Animal Experiments at the Graduate School of Information Science and Technology, the University of Tokyo (Permit Number: JA19-1).

Eight dissociated neuronal cultures at 28–37 days in vitro were used in the experiments. Dissociated neuronal cultures were prepared on a high-density CMOS array as previously reported (33, 48-52). Briefly, cortical tissue was dissected from E18 Wistar rats (Jcl: Wistar, CLEA Japan, Japan) and dissociated in 0.25% trypsin-EDTA solution (Thermo Fisher Scientific Inc., MA, USA) at 37 °C in a thermostatic chamber for 20 min. The dissociated cortical cells were plated on CMOS microelectrode arrays (MEAs) (53) at a concentration of 38 K cells over the recording area of $3.85 \times 2.10$ mm$^2$. The cells were maintained for 24 h in an incubator at 36.5 °C and 5.0% $CO_2$ with plating medium prepared by mixing 450 mL of NeuroBasal (Thermo Fisher Scientific Inc.), 50 mL of horse serum (Cytiva, MA), 10 mL of B27 (Thermo Fisher Scientific Inc.), and 1.25 mL of GlutaMAX (Thermo Fisher Scientific Inc.). Half of the medium in the MEA chambers was replaced with the same amount of growth medium prepared by mixing 450 mL of DMEM (Thermo Fisher Scientific Inc.), 50 mL of horse serum, 5 mL of sodium pyruvate (Thermo Fisher Scientific Inc.), and 1.25 mL of GlutaMAX (Thermo Fisher Scientific Inc.). After plating, the MEAs were kept in an incubator, and half of the medium was replaced with fresh growth medium twice a week throughout the study period.

## 2.9 Action Potential Detection

The signals were filtered using a 300-3000 Hz bandpass. The action potential was detected by crossing the threshold, which was set at –5 times the standard deviation of the measurement noise at each electrode(54), and the time at which the action potential waveform fell below the threshold was recorded as the onset time of the action potential.

## 2.10 Electrode Selection in CMOS MEA

The CMOS MEA was able to simultaneously record neural signals from 1,024 of the 26,400 electrodes. Prior to the main experiments, all 26,400 measurement electrodes were scanned for 1 min to detect action potentials. A maximum of 1,024 electrodes were selected in the order of the negative peak amplitude of the action potentials, and these electrodes were used in the main experiments.

As in the SNN simulations, excitatory neurons were the stimulation targets. Because inhibitory neurons have more potassium channels, Kv3.1 and Kv3.2, than excitatory cells (55, 56), the action potential waveforms from excitatory neurons were different from those from inhibitory neurons in that the time from the negative peak of depolarization to the positive peak of hyperpolarization was longer in excitatory neurons than in inhibitory neurons (51). To identify excitatory neurons, we aligned action potentials with the negative peak of depolarization, calculated the median value of the potential at each time step, and obtained a positive peak. Electrodes with a peak-to-peak time of 0.5 ms or longer were considered below excitatory neurons. Twenty-four electrodes below putative excitatory neurons were randomly selected for stimulation under the condition that these stimulating electrodes were spatially arranged in a circle throughout the entire test area with an inter-electrode distance of 100 μm or more (Fig. 4A(i)). Each pattern consisted of six stimulation electrodes in the vicinity where electric pulses were applied simultaneously (Fig. 4A(ii)). The electric pulse was a positive first biphasic waveform with a width of 200 μs for each phase and an intensity of ±200 mV.

### 2.11 Stimulation to dissociated culture

As in the simulation above, fading memory was characterized in dissociated cultures of neurons. The medium was changed half a day before the start of the experiment and retained during the experiments.

The four stimulation patterns were repetitively applied 900 times, each at 1-second intervals, in a random order, to induce plasticity in neuronal cultures. Decoding accuracy and decay time of fading memory were characterized before the repetitive stimulation and 0.5, 3, 24, and 48 h after the repetitive stimulation from evoked responses to each stimulus were presented 20 times at 17-34 s intervals in a random order.

Because electrical stimulation generates artifacts in the measurements, data within 1 ms of stimulation and data from electrodes saturated by stimulation were excluded from the analyses. After removing artifacts, firing rates were quantified at each measurement electrode with 20-ms bins, and firing rate vectors were constructed at each bin to characterize the classification accuracy and decay time by SLR, as in the SNN simulation.

## 3 Results

### 3.1 Noise-driven Activity Promotes Criticality and EI balance

To examine how noise-driven spontaneous activity might influence criticality and EI balance in a neural network, we designed an SNN comprising 80 excitatory and 20 inhibitory neurons with leaky integrated-and-fire (LIF) dynamics. We implemented short-term plasticity (STP) and STDP in the network. Each neuron was fully connected to all others, with synaptic weights modified by the excitatory and inhibitory STDP rules.

Neuronal firing was stochastic and determined by an exponential function of the difference between the membrane potential and firing threshold. The firing rate at rest was set to 0.4 Hz, introducing noise-driven spontaneous activity into the system. This activity drove the SNN to either a subcritical (criticality index $\Delta Cr<0$), critical ($\Delta Cr \approx 0$), or supercritical ($\Delta Cr > 0$) state, depending on the parameters of the STDP windows.

The excitatory STDP window with the parameter $\beta_E$ was temporally asymmetric, whereas the inhibitory STDP window with $\beta_I$ was symmetric (Fig. 1A). Parameters $\beta_E$ and $\beta_I$ determined the ratio of long-term depression (LTD) to long-term potentiation (LTP) in excitatory and inhibitory synapses, respectively. SNNs with large $\beta_E$ and $\beta_I$ values, that is, LTD more dominant than LTP, became subcritical, whereas SNNs with small $\beta_E$ and $\beta_I$ values became supercritical (Fig. 1B and C). Simulations were run for 72 h with different random seeds to sufficiently stabilize SNNs before the main experiments.

In the critical SNN, raster plots of spontaneous activity revealed neuronal avalanches of various sizes corresponding to the number of simultaneously activated neurons. These avalanches exhibit a power-law size distribution, which is a hallmark of criticality (Fig. 1D (i) Upper left and right). The critical SNN also exhibited EI balance (Fig. 1D (i) lower left). Bursts were driven by simultaneous input from excitatory and inhibitory synapses with a high cross-correlation (>0.75). The subcritical SNN did not show large-scale bursts, and neuronal avalanches were characterized by an exponential size distribution (Fig. 1D (ii)). These firing patterns are predominantly driven by excitatory inputs with virtually no inhibitory inputs. The supercritical SNN exhibited more large-scale bursts with fewer small-scale avalanches than the subcritical SNN, while achieving the same level of EI balance as the critical SNN (Fig. 1D (iii)).

We then examined the robustness of the criticality and EI balance in the SNN when subjected to external stimulation. As a stimulus, we simultaneously delivered external inputs to six excitatory neurons, increasing their membrane potential by 30 mV and eliciting spikes. We repeatedly applied four different test stimuli to the SNN, each of which was applied 900 times at 1 Hz in random order, for a total of 3600 stimuli over one hour.

In the critical SNN, the criticality index ΔCr showed a transient decrease during the repetitive stimulation session and returned to the initial level within two hours, suggesting that criticality was transiently disrupted by repetitive stimulation and subsequently restored (Fig. 2A (i)). Interestingly, in a subcritical SNN, ΔCr tended to show a transient decrease during stimulation, followed by an increase, reaching a stable plateau closer to the critical state than the initial state (Fig. 2A (ii)). Similarly, in a representative supercritical SNN, ΔCr tended to change towards zero after the repetitive stimulation session and a subsequent session of spontaneous activity (Fig. 2A (iii)). This trend was consistently observed within the tested range of the parameters (Fig. 2B); post-stimulus |ΔCr | (12h) was significantly smaller than pre-stimulus |ΔCr | (pre) (Wilcoxon signed rank test, p=1.9e-11). These results support our main hypothesis that spontaneous activity is a driving force towards the critical state, as shown in the development of the SNN, and that repetitive stimulation and subsequent spontaneous activity triggers a transition to a more stable and critical state than the initial state in the SNN.

This transition to criticality is associated with a change in the baseline level of spontaneous activity (Fig. 2C). The baseline levels decreased immediately after repetitive stimulation (pre vs. 0h; Wilcoxon signed rank test (uncorrected): crt, p=1.2e-6; sub, p=1.7e-06; sup, p=1.7e-06). During the subsequent session of spontaneous activity, the baseline level was adjusted; it returned to the initial level in the critical SNN, increased above the initial level in the subcritical SNN (pre vs. 12h, p=6.9e-05), and decreased below the initial level in the supercritical SNN (pre vs. 12h, p=1.7e-06). These results suggest that repetitive stimuli trigger a state transition in the SNN and that the firing rate set point of spontaneous activity changes with the critical state in the SNN.

These changes in criticality and spontaneous activity were also associated with changes in the EI balance and synaptic inputs (Fig. 2D). In the critical SNN, EI balance (cross-correlation between E and I) was transiently disrupted immediately after repetitive stimulation (Wilcoxon signed rank test (uncorrected): pre vs. 0h, p=1.7e-06) but was restored after 12 h of spontaneous activity (0h vs. 12h, p=1.5e-05). Both the E and I inputs showed transient decreases after repetitive stimulation (E, p=1.7e-06 for pre vs. 0h, p=1.7e-06 for 0h vs. 12h; I, p=1.7e-06 for pre vs. 0h, p=1.7e-06 for 0h vs. 12h), whereas the I/E ratio remained constant. In both subcritical and supercritical SNNs, the EI balance changed with repetitive stimulation (sub, p=5.2e-04; sup, p=3.2e-06) and did not return to the original level, whereas E, I, and I/E showed transient changes (E in sub, p=1.7e-06 for pre vs. 0h, p=1.7e-06 for 0h vs. 12h; I in sub, p=1.7e-06 for pre vs. 0h, p=1.7e-06 for 0h vs. 12h; I/E in sub, p=1.7e-06 for pre vs. 0h, p=1.7e-06 for 0h vs. 12h; E in sup, p=1.7e-06 for pre vs. 0h, p=1.7e-06 for 0h vs. 12h; I in sup, p=1.7e-06 for pre vs. 0h, p=3.2e-06 for 0h vs. 12h; I/E in sup, p=2.6e-06 for pre vs. 0h, p=1.7e-06 for 0h vs. 12h). These results are consistent with our hypothesis that the baseline firing rate level, criticality, and EI balance are the set points of SNNs, which are driven by noise and self-organized by a temporally asymmetric excitatory STDP window and a symmetric inhibitory STDP window.

### 3.2 Spontaneous Activity Enhances Stimulus-Specific Memory Through Critical Dynamics

We hypothesized that critical dynamics maintained by spontaneous activity would prolong the fading memory of experienced stimuli. To test this hypothesis, we first quantified the network's ability to maintain stimulus-specific information. Following external stimulation, the SNN showed a transient increase in firing rate with a distinct spatiotemporal pattern and returned to baseline levels of spontaneous firing rate (Fig. 3A and B). To confirm that this transient activity was stimulus specific, we constructed a decoder to predict which of the four test stimuli was presented, based on the spatial firing pattern within a given 20-ms bin. As expected, decoding accuracy decayed over time following the stimulus. To quantify the fading memory strength, we measured the decay time within which the decoding accuracy was above chance level (0.25) (Fig. 3C). The decay time was approximately equal to the time during which the firing rate was temporarily increased by the test stimulus (Fig. 3B and C). The subcritical SNN had a slightly longer decay time of fading memory than either the critical or supercritical SNN (Wilcoxon signed rank test (uncorrected): crt vs. sub, p=0.0014; crt vs. sup, p=0.062; sub vs. sup, p=1.9e-5). The decay time was consistent across the different test bins, ranging between 5 and 100 ms (Supplementary Fig. S1).

Next, we reasoned that the evoked responses and fading memory would change after repeated stimulation and subsequent spontaneous activity. We repeatedly applied the four test stimuli to the SNN, each of which was applied 900 times at 1 Hz in random order, and examined how repeated stimulation and subsequent spontaneous activity affected fading memory. After 12 h of spontaneous activity following the stimulation session, we found that evoked responses were enhanced with longer decay times of fading memory (Fig. 3C and D) (Wilcoxon signed rank test (uncorrected) for pre vs. 12h: crt, p=1.7e-6; sub, p=1.7e-6; sup, p=8.4e-3). This enhancement was most pronounced in the critical SNN and virtually negligible in the supercritical SNN. Interestingly, although the fading memory of the first stimulation lasted longer in the subcritical SNN than in the critical SNN, the fading memory of the repeated stimulation became longer in the critical SNN than in the subcritical SNN after the 12-h spontaneous activity session (Wilcoxon signed rank test (uncorrected), p=4.1e-6), suggesting that criticality possibly maintained by spontaneous activity plays an important role in this type of memory consolidation.

To confirm the functional role of spontaneous activity, we examined how the stimulus-evoked response and fading memory in the critical SNN changed with spontaneous activity after repetitive stimulation sessions. We found that the evoked firing rates gradually increased over the first hour and then rapidly escalated in a step-like manner (Fig. 3E). The decay time of the fading memory exhibited a similar pattern (Fig. 3F; Wilcoxon signed rank test (uncorrected): pre vs. 12h, p=1.7e-06). This spontaneous activity-driven prolongation of fading memory was observed at an adequate level of spontaneous activity (Supplementary Fig. 2A). In contrast, the subcritical SNN exhibited more gradual increases in evoked firing rates and fading memory without the abrupt transitions seen in the critical SNN, whereas the supercritical SNN showed minimal changes (Supplementary Fig. S3). We also confirmed that these changes were dependent on the number of stimulus repetitions during the stimulation session (Fig. 3G). Both the evoked responses and the decay time of the fading memory were expressed as a sigmoid-like function of the number of repetitions (Fig. 3H; Kruskal-Wallis test, df=179, Chi-sq=93.5, p=1.2e-18). These findings indicate that both the plasticity directly induced in the repetitive stimulation session and the subsequent self-organization during spontaneous activity play important roles in enhancing evoked responses to repeated stimuli and prolonging fading memory. Similar to the change in criticality and baseline firing rate, a sharp, step-like change in evoked responses and fading memory during spontaneous activity after a repetitive stimulation session could represent a state transition.

Long-term repetitive stimulation-induced changes in evoked responses and fading memory are likely to be driven by excitatory synaptic plasticity. Following repetitive stimulation, 6% of $w_{EE}$ and 2% of $w_{EI}$ in the critical SNN increased from near zero to near one, of which 85% and 58%, respectively, were maintained at 0.5 or higher for 12 h (Supplementary Fig. S4); and in the subcritical SNN, 9% of the $w_{EE}$ and 10% of the $w_{EI}$ increased, of which 91% and 85% were maintained for 12 h (Supplementary Fig. S5). The plasticity of inhibitory synapses was transient, suggesting that inhibitory synapses play a modulatory role: 28% of $w_{IE}$ in the critical SNN, 79% of $w_{IE}$, and 94% of $w_{II}$ in the subcritical SNN were strengthened by repetitive stimulation, but almost none of them was maintained for 12 h (Supplementary Fig. S4 and S5). The synaptic plasticity in the supercritical SNN was less distinct than that in the critical and subcritical SNNs (Supplementary Fig. S6): Repetitive stimulation strengthened 5% of $w_{EE}$, 98% of which was maintained for 12 h, but virtually no plasticity was observed in $w_{EI}$, $w_{IE}$, and $w_{II}$. This synaptic strengthening and maintenance also depended on the noise level; at high noise levels, excitatory synaptic weights were more likely to be temporarily strengthened but less likely to be maintained in the long term, whereas inhibitory synaptic weights were more likely to be maintained in the long term (Supplementary Fig. S2B and C).

### 3.3 Spontaneous Activity Enhances Stimulus-Specific Memory in Dissociated Neuronal Cultures

To validate our computational predictions in a biological context, we examined whether fading memory was prolonged by spontaneous activity after repeated stimulation of test patterns in dissociated neuronal cultures. Previous studies have shown that these cultures exhibit stimulus-specific plasticity(34-37), but have overlooked the functional role of spontaneous activity in this type of learning.

We recorded spatiotemporal neural activity using a high-density CMOS microelectrode array (33, 50, 57, 58) (MaxOne High-Density Microelectrode Array System, MaxWell Biosystems AG, Switzerland). Of 26,400 sites, 1,024 electrodes were selected for recording in the order of the negative peak amplitude of the action potentials (Supplementary Fig. S7) and 24 for stimulation (Fig. 4A (i)). Of the 24 stimulation electrodes arranged in a circle in the test area, we grouped six neighboring electrodes to create four distinct stimulus patterns (Fig. 4A (ii)). Each stimulus pattern was delivered 900 times at 1 Hz in random order, replicating the protocol used in our SNN simulations.

Consistent with our computational findings, we examined the fading memory of test patterns before and after stimulation with spontaneous activity. To quantify fading memory, the four stimulus patterns were delivered 20 times at 17-34 s intervals in a random order, and the stimulus-evoked responses were measured. Figure 4B shows representative raster plots of stimulus-evoked activity before and 24 h after stimulation sessions. Repetitive stimulation and 24 h of spontaneous activity were effective in increasing the stimulus-evoked activity in some electrodes. Associated with this stimulus-specific plasticity, the decoding accuracy of the test stimulus also gradually increased with the spontaneous activity session, and was maximized 24 h after the stimulation session (chance level = 0.25) (Fig. 4C). Consequently, as in the SNN simulation, we confirmed that the decay time of the fading memory increased significantly 24 h after the stimulation session compared to the pre-stimulation session (Fig. 4D; p=0.038, Wilcoxon signed-rank test). In addition to the decoding-based analysis, information theoretical analysis estimating mutual information between the firing rates and the stimulus pattern also supports that spontaneous activity following repetitive stimulation makes the evoked response more informative (Supplementary Fig. S8) (59). During the measurements, we did not observe any significant changes in the baseline levels of spontaneous activity (Fig. 4E). These results were consistent with the SNN simulation, supporting our claim that spontaneous activity after the stimulation session has a comparable baseline level to the pre-stimulation session and is crucial for strengthening memory of past inputs without inducing overall activation of neuronal culture. On the other hand, the results are partially inconsistent with the SNN simulation in that the stimulation session did not induce a transient change in the baseline firing rate.

## 4 Discussion

Both experiments in the SNN and dissociated culture of neurons showed that spontaneous activity played a crucial role in maintaining the critical state and EI balance and prolonging the fading memory of stimuli that have been experienced repeatedly in the past, while repeated stimuli transiently disrupted criticality. These results support our hypothesis that the criticality and EI balance in the neural network are homeostatically maintained by spontaneous activity with appropriate noise levels and STDP windows and play an important role in the consolidation of memory for past experiences.

### 4.1 Homeostatic set points emerging in SNN with STDP

State transitions were also observed during SoC development. The developmental stages of the neuronal culture were classified by the states of criticality(33, 60-65): (i) “subcritical state” in immature culture exhibiting asynchronous activity pattern with an exponential size distribution of neuronal avalanche; (ii) “supercritical state” in developing culture characterized as regularly synchronized bursts and a bimodal distribution; and (iii) “critical state” exhibiting diverse size of avalanches with a power-law distribution. The mechanism underlying these state transitions is analogous to that observed in the stimulus-induced reorganization of the network investigated in this study.

In our model, noise-driven stochastic discharges are the driving force of SoC(11). Under noise, excitatory STDP with temporally asymmetric windows dichotomizes strong versus weak synapses and high- versus low-firing neurons, thereby destabilizing the network towards the supercritical, synchronized state. Subsequently, inhibition stabilizes the synchronized state towards the critical state and balances excitation because synchronized bursts are more effective on temporally symmetric inhibitory STDP windows than on asymmetric excitatory STDP windows. Thus, inhibition stabilization is likely the key mechanism of SoC(41), in addition to efficient coding and nonlinear computation, as discussed in the past (15-20). This study further emphasizes the functional role of spontaneous activity and STDP windows in the regulation of inhibitory tone.

In simulation of spontaneously developed SNNs, EI balance (high CC between E and I inputs) was observed in the critical state and supercritical state (CC≒0.9), but not in the subcritical state (CC<0.5) (Fig. 1B). The subcritical state had fewer synaptic inputs than the critical and supercritical states did. Furthermore, the I/E ratio in the subcritical state was close to zero, indicating that virtually no inhibitory synapses were formed. Because the symmetric STDP window of inhibitory synapses is effective for burst activities for LTP induction, subcritical states with fewer bursts were unable to grow more inhibitory synapses than the critical and supercritical states. The supercritical state achieved an EI balance but exhibited less diversity of burst sizes than the critical state, according to the definition of the neuronal avalanche distribution. The bursts in the supercritical state were more homogeneous than those in the critical state because the excitatory input strengths varied less. Furthermore, bursts in the supercritical state tended to terminate more quickly than those in the critical state because of the higher I/E ratio; this trend was more obvious in stimulus-evoked activity (Fig. 2A).

The test indices of synaptic activity in Fig. 2D depend on the noise level and the STDP window. In all of these networks, the test indices transiently changed in response to repeated external stimuli (pre vs. 0 h) but tended to return to their original state after 12-h of spontaneous activity (pre vs. 12 h). The baseline level of the spontaneous firing rate also transiently decreased immediately after the repeated stimuli but tended to return to the original level after spontaneous activity (Fig. 2D). These results suggest that the baseline level of synaptic activity and spontaneous firing rate depend on the noise level and the STDP window. Interestingly, however, after repeated stimuli and subsequent spontaneous activity sessions, the baseline level of spontaneous firing rate slightly increased in the subcritical state and decreased in the supercritical state. Associated with these baseline shifts (Fig. 2D), the subcritical and supercritical networks become more critical than the original states (Fig. 2A and B). Stimulus-induced plasticity likely caused the non-critical networks to escape from their locally stable states, and the subsequent spontaneous activity brought the networks closer to the critical state. Thus, even under suboptimal conditions, the noise and STDP in our SNN configuration act as pressures that drive neural networks to a critical state.

### 4.2 Stimulus-specific fading memory

Both the SNN and dissociated culture of neurons showed that the decay time of the fading memory of stimuli that were repeatedly exposed and consolidated through sufficient spontaneous activity became longer in the critical state than in either the subcritical or supercritical states (Fig. 3 and 4). Furthermore, the SNN simulation showed that the more often a stimulus was repeated, the stronger their fading memory (Fig. 3G and H), suggesting that fading memory depended on the direct modification of the network topology during the stimulation session rather than the subsequent spontaneous activity. However, this lengthening of fading memory was less obvious in the supercritical state because the stimulus-evoked activity ended up in a burst, which was quickly terminated by strong inhibition (Fig. 3A and B). These results support our hypothesis that the homeostatic set point of criticality has the advantage of encoding input stimuli experienced in the past in neural networks.

The decay times of stimulus-evoked firing rate (Fig. 3A and B) did not always behave as those of decoding accuracy (Fig. 3C); for example, before repeated stimulation, the firing rates in the critical state lasted longer than those in the subcritical state, but the decoding accuracy in the critical state decayed faster than that in the subcritical state. Furthermore, prolonged fading memory, or transient activity increase and its decay, after repeated stimulation was stimulus specific. These observations suggest that the decoding accuracy is not directly related to a general (i.e., stimulus-unspecific, network-intrinsic) time constant of neural dynamics, e.g., autocorrelation times.

Unintuitively, fading memory before repeated stimulation was longer in the subcritical state than in the critical state (Fig. 2C and D). The subcritical state, characterized by weak coupling between excitatory and inhibitory inputs, produces a highly redundant response from different neurons in response to external stimuli (66). Such evoked responses in the subcritical state are likely to be more readily decodable than those in the critical state, but in turn, may limit complex information propagation. Furthermore, considering the trade-off between the sensitivity and specificity of neural representation, the subcritical SNN may better represent the stimulus than the critical SNN when the stimulus is sufficiently distinct (67).

The simulation in the SNN and physiological experiments in dissociated cultures of neurons differed in several respects. First, as predicted by our hypothesis, repeated stimulation temporarily transitioned the critical SNN to the subcritical state with a reduced baseline firing rate, and subsequent spontaneous activity restored the critical state. However, such two-step state transitions were not observed in neuronal cultures. Second, in neuronal culture, the decay time of the fading memory was lengthened after 24-h of spontaneous activity but returned to the original level after 48 h, while the fading memory in the SNN remained high for the entire test period. Further experiments both in silico and in vitro are needed to determine whether our SNN missed a key parameter or whether large variations between neuronal cultures in our physiological experiments caused limited agreement between in silico and in vitro experiments.

### 4.3 Limitation

Our model was effective in simultaneously capturing the essential characteristics of the SoC and memory consolidation but ignored many important aspects of the brain. For example, the diversity of ion channels, such as slow NMDA and $GABA_B$ receptors, might produce richer dynamics and plasticity than our simple model (68-70)(71). Although our model consists of only two types of neurons, excitatory and inhibitory neurons, the cerebral cortex in mice consists of at least 41 types of excitatory neurons and 34 types of inhibitory neurons(72) with distinct motifs of local microcircuits(73-75). STDP windows depend on the cell type and dendritic location(76, 77) under the control of neuromodulation and glial activity(78-80). The EI balance is controlled not only by STDP but also by transcellular chemical signaling (81, 82). Further studies are required to elucidate how these factors contribute to the development of SoC and memory consolidation.

### 4.4 Physiological implications

We propose that the neural mechanisms discussed here may be implemented in the brain to maintain criticality during sleep because criticality is likely to be restored during sleep and progressively disrupted during waking experience(42, 83, 84), as we demonstrated that criticality is restored by spontaneous activity and disrupted by external stimuli. Sleep could be the price that the brain has to pay for plasticity during the waking experience(85); therefore, sleep pressure increases when the brain deviates from the critical state, indicated by slow wave activity (86, 87), intermittent activity (88), low complexity of activity (89), activity level (90, 91), synaptic scaling (92-96) and EI balance (97), many of which could be explained by criticality. Our model suggests that the brain reestablishes a critical state through a push-pull mechanism between the subcritical and supercritical states when deviating from the critical state. This mechanism is different from prior modeling, suggesting that sleep plays an essential role to make a margin from the supercritical state (98). In addition to maintaining criticality as a homeostatic property, sleep is also critical for memory consolidation and integration(99-101), which is consistent with our finding that spontaneous activity after repeated stimuli prolongs stimulus-specific fading memory. We speculate that oscillation during sleep (e.g., ripples and spindles) has the same function as bursts in our configuration, which selectively strengthens inhibitory synapses and develop an EI balance (87, 102). Unlike our stimulation experiments, the firing rates sometimes increased by learning (cf. (103)). In such case, our observation in the network development simulation predicts that the increase of firing rates is associated with transient increase of burst activity, which in turn, developed the inhibitory synapses due to the temporally symmetric STDP window and EI balance (i.e., inhibition stabilization).

## 5 Conclusion

Our findings suggest that the brain may achieve self-optimization and memory consolidation as emergent functions of noise-driven, spontaneous activity. Noise-harnessing computation represents an evolutionary adaptation of the brain (104), which has been destined to be as energy-efficient as possible and to operate in harsh biochemical environments with low signal-to-noise ratios. Other examples of emergent neural mechanisms utilizing noise include stochastic resonance (105, 106), simulated annealing (107) and noise-induced chaos-order transitions (108, 109). In contrast, a standard digital computer requires more energy than the brain because of the presence of a significant margin between high and low levels, that is, 0 V vs. 5 V, which is necessary to maintain an adequate signal-to-noise ratio (110-112). This evolutionary trait in the brain offers valuable insight into the design principles of energy-efficient large-scale SNNs.

## Figures

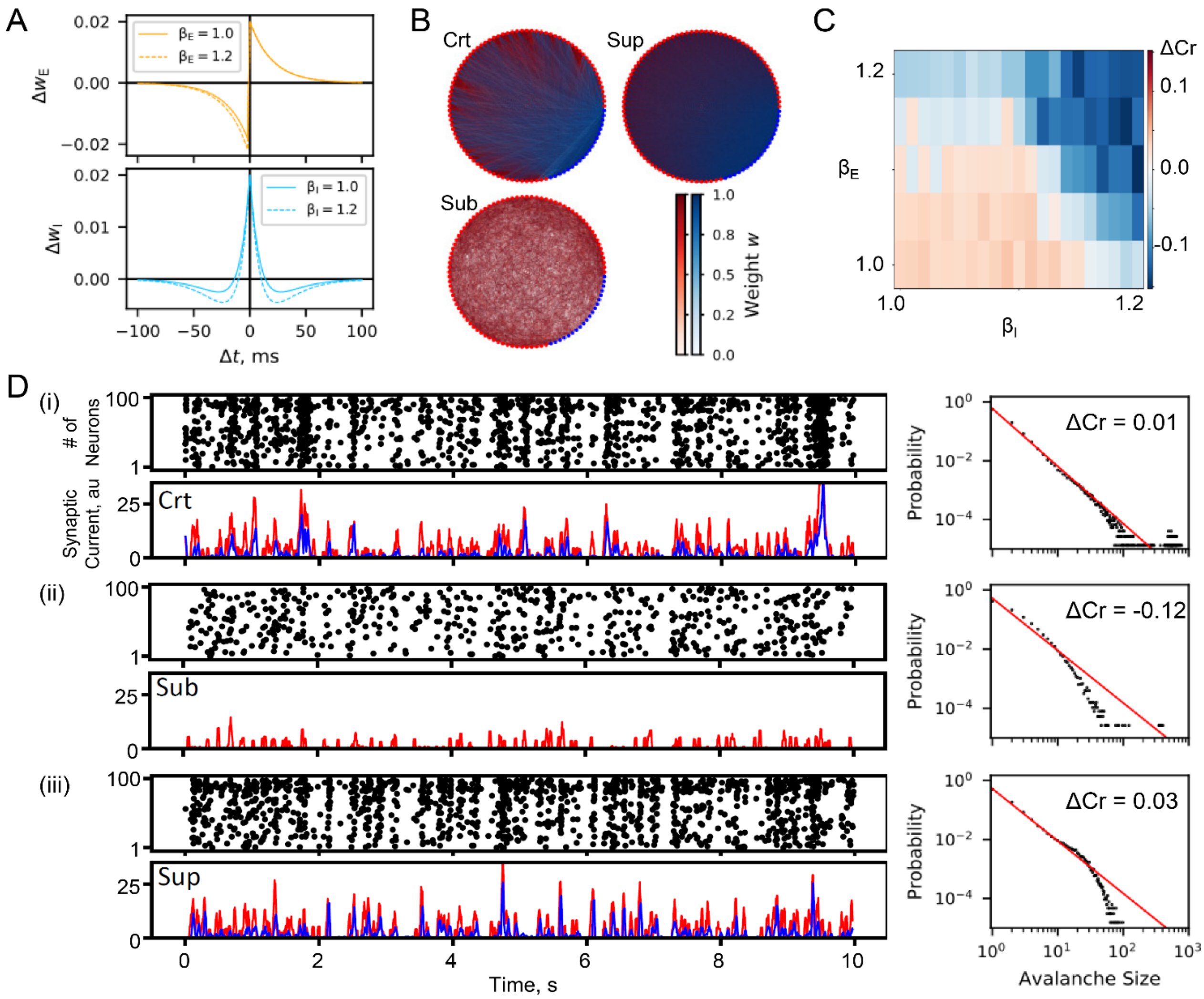

**Figure 1.** SNN simulation. (A) STDP windows for excitatory ($\Delta w_E$) and inhibitory synapses ($\Delta w_I$). Parameters $\beta_E$ and $\beta_I$ determine the balance between LTP and LTD in excitatory and inhibitory neurons, respectively. (B) Representative networks in the critical, subcritical, and supercritical states (Crt, Sub, and Sup). The dots arranged in a circle represent neurons: red, excitatory; and blue, inhibitory. The thickness of the line between neurons represents the synaptic weights. (C) The criticality index (ΔCr) as a function of $\beta_E$ and $\beta_I$. (D) Representative results for (i) critical, (ii) subcritical, and (iii) supercritical SNNs. For each SNN, the upper left, lower left, and right insets show a raster plot, synaptic currents (red, excitatory input; blue, inhibitory input), and neuronal avalanche size distribution, respectively.

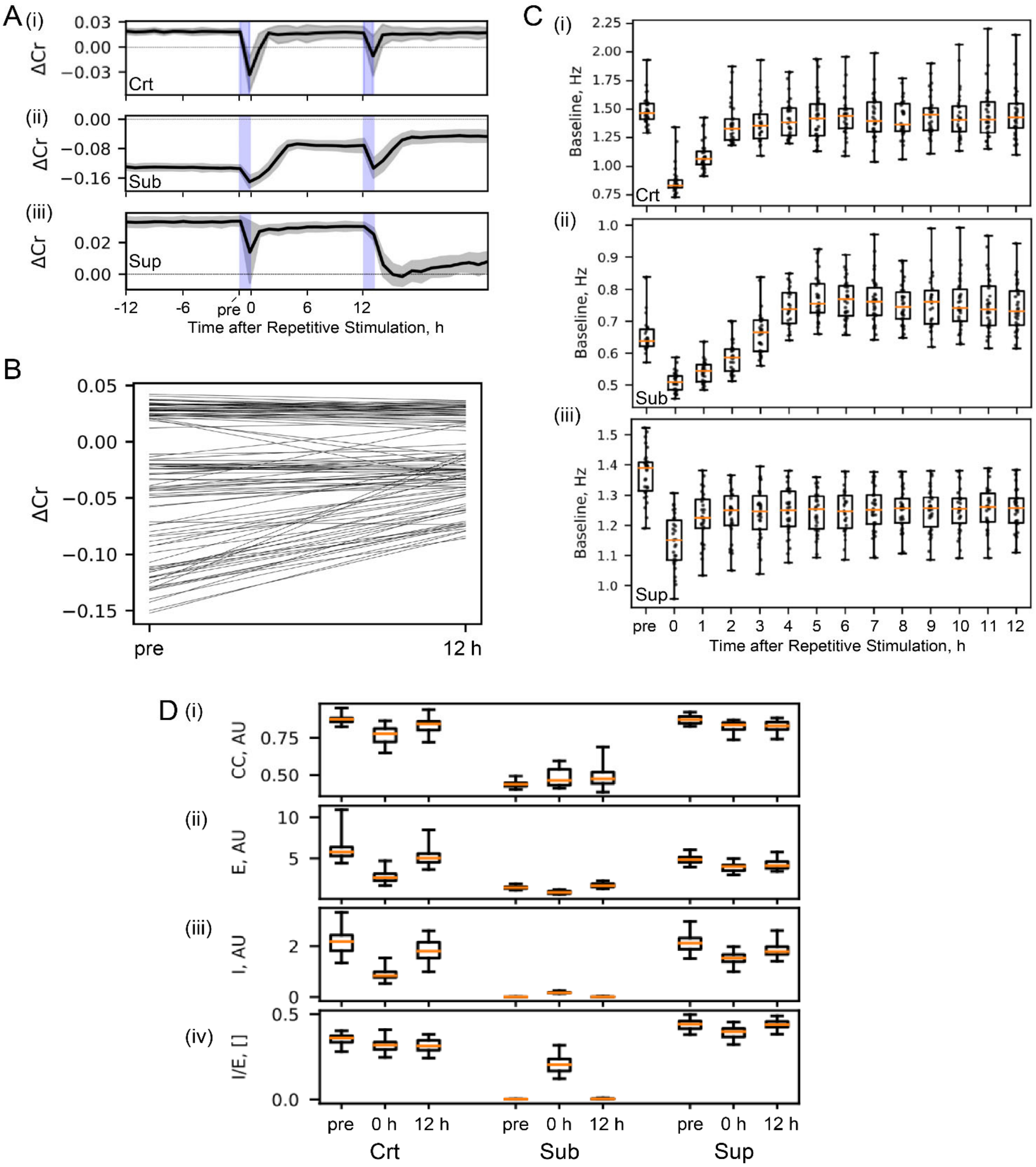

**Figure 2.** Fig. 2 Homeostatic set points in the SNN. (A) Criticality. The mean and standard deviation of ΔCr from different initial seeds (n=30) are shown in representative critical, subcritical, and supercritical SNNs (Crt, $\beta_E$=1 and $\beta_I$=1.15; Sub, $\beta_E$=1.2, and $\beta_I$=1.2; Sup, $\beta_E$=1 and $\beta_I$=1). Repetitive stimulation for one hour was performed twice (shaded areas in blue). (B) Stimulation-triggered criticality transitions pre- and post-stimulation (with 12-h spontaneous activity session) ΔCr were compared in different SNNs with $\beta_E$ and $\beta_I$, as shown in Fig. 1B. Post-stimulus |ΔCr | (12h) was significantly smaller than pre-stimulus |ΔCr | (pre) (see the main text for statistics). (C) Baseline spontaneous firing rate. Data from the representative (i) critical, (ii) subcritical, and (iii) supercritical

SNNs are shown. In each boxplot, the central mark indicates the median and the upper and lower edges of the box indicate the 75th and 25th percentiles, respectively; whiskers extend to the most extreme data points. The baseline levels at 0h were significantly smaller than those at pre in all of the SNN (see the main text for statistics). The post-stimulus baseline levels increased above the pre-stimulus level in the subcritical SNN, and decreased below the pre-stimulus level in the supercritical SNN (see the main text for statistics). (D) EI balance. Pre-stimulus, immediately after post-stimulus, and 12 h after post-stimulus sessions were characterized (pre, 0, and 12) for the critical, subcritical, and supercritical SNNs (Crt, Sub, and Sup). (i) The cross-correlation (CC) of excitatory and inhibitory inputs is shown in critical, subcritical, and supercritical SNNs. (ii) and (iii) Mean amplitudes of excitatory and inhibitory inputs (E and I). (iv) Ratio of I-to-E magnitude. The changes in criticality and spontaneous activity (A-C) were associated with the changes in these parameters (see the main text for statistics).

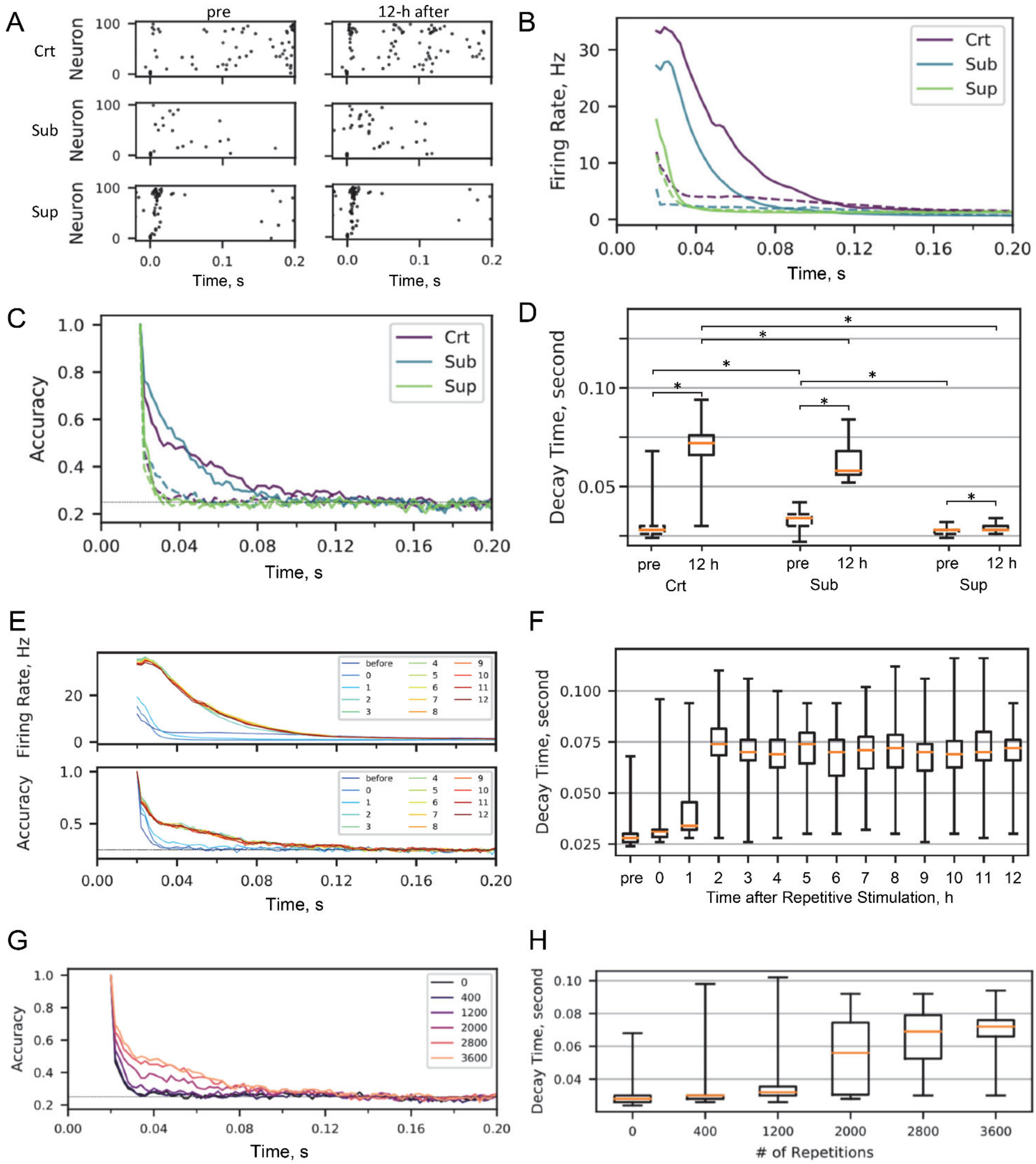

**Figure 3.** Fading memory in SNN. (A) Representative evoked responses. Raster plots are shown before and 12 h after repetitive stimulation sessions in critical, subcritical, and supercritical SNNs. (B) Traces of the firing rate after stimuli: dashed line, before repetitive stimulation; solid line, 12 h after repetitive stimulation. (C) Decoding accuracy of the test stimulus from a spatial activity pattern within a given 20-ms bin. (D) Decay time of fading memory before and 12 h after repetitive stimulation. Asterisks indicate statistical significance (see the main text for statistics). (E) Traces of firing rate and decoding accuracy at the indicated times after repetitive stimulation. (F) Traces of the fading memory decay time after repetitive stimulation. (G) Decoding accuracy as a function of number of stimulus repetitions. (H) Decay time of fading memory as a function of the number of stimulus repetitions.

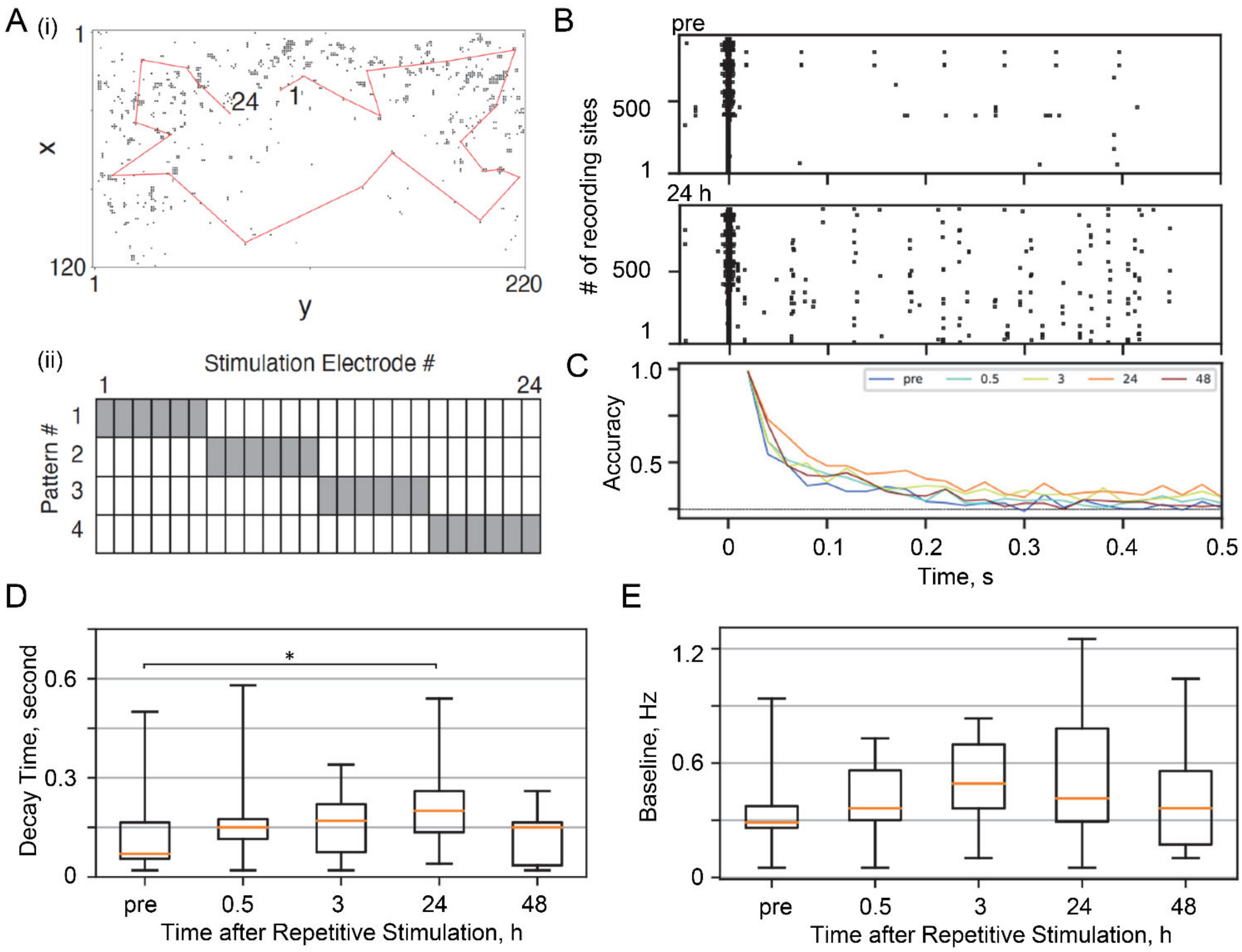

**Figure 4.** Fig. 4 Physiological validation in dissociated neuronal cultures. (A) Representative configuration of the neurons in the microelectrode array. (i) Recording was performed at 1024 of 26400 sites (dots), and stimulation was provided at 24 sites on a red line. These electrodes were selected in the order of the negative peak amplitude of the action potentials (Supplementary Fig. S7). (ii) For stimulation patterns, six sites in the vicinity were combined. (B) Representative evoked responses at each recording site. Raster plots were obtained before and 24 h after the repetitive stimulation sessions. (C) Decoding accuracy at the indicated times after repetitive stimulation. (D) Fading memory decay time after repetitive stimulation. Asterisks indicate statistical significance (see the main text for statistics). (E) Baseline spontaneous firing rate.

**Tables**

**Table 1.** Spiking neural network parameters

| Parameter | Value |
|---|---|
| $\Delta t$ | 0.1 ms |
| $\tau_m$ | 30 ms |
| $v_{rest}$ | -74 mV |
| $v_{th}$ | -54 mV |
| $E_{exc}$ | 0 mV |
| $E_{inh}$ | -80 mV |
| $f_{rest}$ | 0.4 Hz |
| $b$ | 4 mV |
| $T_{exc}^{ref}$ | 3 ms |
| $T_{inh}^{ref}$ | 2 ms |
| $\tau_{AMPA}$ | 2 ms |
| $\tau_{GABA}$ | 4 ms |
| $\tau_{rec}$ | 150 ms |
| $U$ | 0.4 |
| $g_{exc}^{max}$ | 4.0 |
| $g_{inh}^{max}$ | 4.0 |
| $A_E$ | 0.02 |
| $A_I$ | 0.02 |
| $\tau_E$ | 20 ms |
| $\tau_{I1}$ | 10 ms |
| $\tau_{I2}$ | 20 ms |

**Ethics statement**

The experimental protocol was approved by the Committee on the Ethics of Animal Experiments at the Graduate School of Information Science and Technology, the University of Tokyo (Permit Number: JA19-1). All experimental procedures were carried out in accordance with these approved guidelines.

**Conflict of Interest**

The authors declare that they have no conflicts of interest.

**Author Contributions**

NI: Methodology, Software, Validation, Formal analysis, Investigation, Data Curation, Writing - Original Draft, Visualization; DA: Software, Validation, Formal analysis, Investigation, Data Curation, Writing - Review & Editing, Visualization; HT: Conceptualization, Methodology, Validation, Resources, Writing - Review & Editing, Visualization, Supervision, Project administration.

**Funding**

This work is partly supported by JSPS KAKENHI (23H03465, 24H01544, 24K20854, 25H02600), AMED (24wm0625401h0001), JST (JPMJPR22S8), the Asahi Glass Foundation, and the Secom Science and Technology Foundation.

**Data Availability Statement**

The raw data supporting the conclusions of this article will be made available by the authors, without undue reservation.